\documentclass[aps, prl,twocolumn,superscriptaddress,showpacs]{revtex4}

\usepackage{epsfig}

\usepackage{amsmath,amssymb}

\begin{document}

\title{Entangling flux qubits with a bipolar dynamic inductance}

\author{B. L. T. Plourde}
\affiliation{Department of Physics, University of California,
Berkeley, CA 94720 }

\author{J. Zhang}
\affiliation{Department of Electrical Engineering and Computer
  Sciences, University of California, Berkeley, CA 94720}
\affiliation{Department of Chemistry, University of California,
Berkeley, CA 94720 }

\author{K. B. Whaley}
\affiliation{Department of Chemistry, University of California,
Berkeley, CA 94720 }

\author{F. K.\ Wilhelm}
\affiliation{Sektion Physik and CeNS, Ludwig-Maximilians-Universit\"at,
Theresienstr.\ 37, 80333 M\"unchen, Germany}

\author{T. L. Robertson}
\affiliation{Department of Physics, University of California,
Berkeley, CA 94720 }

\author{T. Hime}
\affiliation{Department of Physics, University of California,
Berkeley, CA 94720 }

\author{S. Linzen}
\affiliation{Department of Physics, University of California,
Berkeley, CA 94720 }

\author{P. A. Reichardt}
\affiliation{Department of Physics, University of California,
Berkeley, CA 94720 }

\author{C.-E. Wu}
\affiliation{Department of Physics, University of California,
Berkeley, CA 94720 }

\author{John Clarke}
\affiliation{Department of Physics, University of California,
Berkeley, CA 94720 }

\date{\today}  

\begin{abstract}
  We propose a scheme to implement variable coupling between two flux
  qubits using the screening current response of a dc Superconducting QUantum
  Interference Device (SQUID).  The coupling strength is adjusted by
  the current bias applied to the SQUID and can be varied
  continuously from positive to negative values,
  allowing cancellation of the direct mutual inductance
  between the qubits.
  We show that this variable coupling scheme permits efficient
  realization of universal quantum logic.
  The same SQUID can be used to determine the flux states of the qubits.
\end{abstract}

\maketitle
A rich variety of quantum bits (qubits) is being explored for possible
implementation in a future quantum computer \cite{roadmap}. Of these,
solid state qubits are attractive because of their inherent
scalability using well established microfabrication techniques. A
subset of these qubits is superconducting, and includes devices based
on charge \cite{Nakamura99,Vion02}, magnetic flux
\cite{Orlando99,Friedman00,Wal00}, and the phase difference
\cite{Martinis02} across a Josephson junction.
To implement a quantum
algorithm, one must be able to entangle multiple qubits, so that
an interaction term is required in the Hamiltonian describing a two qubit
system.  For two superconducting flux qubits, the natural interaction
is between the magnetic fluxes. Placing the two
qubits in proximity provides a permanent coupling through their mutual
inductance~\cite{Majer03}. Pulse sequences for generating entanglement
have been derived for several superconducting qubits with
fixed interaction energies \cite{Yamamoto03,Strauch03}. However,
entangling operations can be much more efficient if the interaction
can be varied and, ideally, turned off during parts of the manipulation.
A variable coupling scheme for charge-based superconducting qubits
with a bipolar interaction has
been suggested recently \cite{Averin03}. For flux qubits,
while switchable couplings have been proposed previously
\cite{Mooij99,Clarke02},
these approaches 
do not enable one to turn off the coupling entirely and require
separate coupling and flux readout devices.

In this Letter, we propose a new coupling scheme
for flux qubits in which the interaction is adjusted by changing a relatively
small current. 
For suitable device parameters the sign of the coupling can also be
changed, thus making it possible to null out the direct
interaction between the flux qubits.
Furthermore, the same device can
be used both to vary the coupling and to read out the flux states of
the qubits.
We show explicitly how this variable 
qubit coupling can be combined with
microwave pulses to perform the quantum Controlled-NOT (CNOT) logic gate.
Using microwave
pulses also for arbitrary single-qubit operations,
this scheme provides all the necessary ingredients for
implementation of scalable univeral quantum logic.

\begin{figure}
\centering
\includegraphics{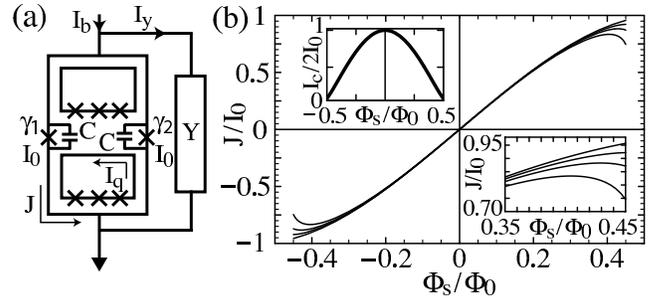}
\caption{(a) SQUID-based coupling scheme. The admittance Y represents
  the SQUID bias circuitry. (b)
  Response of SQUID circulating current $J$ to applied flux $\Phi_s$
  for $\beta_L=0.092$ and $I_b/I_c(0.45 \Phi_0)=0,0.4,0.6,0.85$ (top
  to bottom). Lower right inset shows $J(\Phi_s)$ for same values 
  of $I_b$ near $\Phi_s=0.45 \Phi_0$. 
  Upper left inset shows $I_c$ versus $\Phi_s$.}
\label{layout}
\end{figure}

The coupling is mediated by the circulating current
$J$ in a dc Superconducting QUantum Interference
Device (SQUID), in the zero voltage state, which is coupled to each 
of two qubits through an
identical mutual inductance $M_{qs}$ [Fig. \ref{layout}(a)].
A variation in the flux applied to the SQUID, $\Phi_s$, changes $J$
[Fig. \ref{layout}(b)]. The response is governed by the
screening parameter $\beta_L \equiv 2 L I_0 / \Phi_0$ and the bias current
$I_b$, where $I_b < I_c(\Phi_s)$, the critical current for
which the SQUID switches out of the zero voltage state at $T=0$ in
the absence of quantum tunneling.
In flux qubit experiments \cite{Chiorescu03}, 
the flux state is determined by 
a dc SQUID to which fast pulses of $I_b$ are applied to measure 
$I_c(\Phi_s,T)$. Thus,
existing technology allows $I_b$ to be varied rapidly,
and a single dc SQUID can be used both to measure the 
two qubits and to couple them together controllably.

The flux qubit consists of a superconducting loop interrupted by
three
Josephson tunnel junctions \cite{Mooij99,Orlando99}. With a flux bias near 
the degeneracy point, $\Phi_0/2$, a screening
current $I_q$ can flow in either direction around the qubit loop. Given the
tunnel coupling energy $\delta$ between the different directions of $I_q$, the
ground and first excited states of the qubit correspond to symmetric and
antisymmetric superpositions of these two current states. Thus, the
dynamics of qubit $i$ can be approximated by the two-state Hamiltonian
\begin{equation}
\mathcal{H}_i=-(\epsilon_i^0/2)\sigma_z^{(i)}-(\delta_i/2)\sigma_x^{(i)}.
\label{eq:single_qubit}
\end{equation} 
The energy biases $\epsilon_i^0$ are determined by the flux
bias of each qubit relative to $\Phi_0/2$. The tunnel frequencies $\delta_i/h$ 
are fixed by the 
device parameters,
and are typically a few GHz. 
For two flux qubits, arranged so that a flux change in one qubit 
alters the flux in the other, the coupled-qubit Hamiltonian 
describing the dynamics in the complex 4-dimensional Hilbert space becomes
\begin{equation}
\mathcal{H}=\mathcal{H}_1\otimes I^{(2)} + I^{(1)} \otimes
\mathcal{H}_2
-(K/2) \sigma_z^{(1)}\otimes \sigma_z^{(2)}, 
\label{eq:hamiltonian}
\end{equation}
where $I^{(i)}$ is the identity matrix for qubit $i$ and $K$
characterizes the coupling energy. For $K<0$, the minimum 
energy configuration corresponds to anti-parallel fluxes.
For two flux qubits coupled through a mutual
inductance $M_{qq}$, the interaction energy is fixed at 
$K_0=-2M_{qq}\left|I_{q}^{(1)}\right|\left| I_{q}^{(2)}\right|.$

For the configuration of Fig. \ref{layout}(a), in addition to
the direct coupling, $K_0$, the qubits interact by changing the
current $J$ in the
SQUID. The response of $J$ to a flux change
depends strongly on $I_b$
[Fig. \ref{layout}(b)]. 
When $I_q^{(2)}$ switches direction, the flux coupled to
the SQUID, $\Delta \Phi_s^{(2)}$, induces a change $\Delta J$ in the circulating
current in the SQUID, and alters the flux coupled from
the SQUID to qubit 1. 
The corresponding coupling is
\begin{eqnarray}
K_s = I_q^{(1)}\Delta \Phi_q^{(1)} 
= -2 M_{qs}^2 \left|I_{q}^{(1)}\right|\left| I_{q}^{(2)}\right| 
{\rm Re}\left(\partial J/\partial \Phi_{s}\right)_{I_b}.\label{eq:coupling}
\end{eqnarray}
The transfer function, $\left(\partial J/ \partial \Phi_s\right)_{I_b}$, is 
related to
the dynamic impedance, ${\cal Z}$, of the SQUID via \cite{Hilbert85}
\begin{eqnarray}
\partial J / \partial \Phi_s = i \omega / {\cal
  Z} = 1/{\cal L}+i \omega/{\cal R},
\end{eqnarray}
where ${\cal R}$ is the dynamic
resistance, determined by $Y$ which dominates any loss in the
Josephson junctions, and ${\cal L}$ is the dynamic inductance which, 
in general, differs from the geometrical inductance of the SQUID, $L$. 

We evaluate $\left(\partial J/ \partial \Phi_s\right)_{I_b}$
by current conservation, 
neglecting currents flowing through the junction resistances:
\begin{eqnarray}
I_b&=&I_y+2I_0\cos\Delta\gamma\sin\bar{\gamma}
-2C(\Phi_0/2\pi)\ddot{\bar\gamma},\label{eq:biascurrent}\\
J&=&I_0\cos\bar{\gamma}\sin\Delta\gamma-C(\Phi_0/2\pi)\Delta\ddot{\gamma}.
\label{eq:screeningcurrent}
\end{eqnarray}
Here, $I_y$ is the 
current flowing through the admittance $Y(\omega)$ [Fig.
\ref{layout}(a)], and $I_0$ and $C$ are the critical current and capacitance
of each SQUID junction.  The
phase variables are related to the phases across each junction, $\gamma_1$ and
$\gamma_2$, as
$\Delta\gamma=(\gamma_1-\gamma_2)/2$ and
$\bar{\gamma}=(\gamma_1+\gamma_2)/2$. 
The phases are constrained by
$d\Delta\gamma=(\pi/\Phi_0)(d\Phi_s-LdJ).$

The expression for $K_s$ in terms of ${\rm Re}\left(\partial
  J/\partial \Phi_{s}\right)_{I_b}$ (Eq.~\ref{eq:coupling}) requires the
qubit frequencies to be much lower than the
characteristic frequencies of the SQUID. This condition is satisfied
by our choice of device parameters, and also ensures that the SQUID stays
in its ground state during qubit entangling
operations. Furthermore, it is a reasonable approximation to take the
$\omega=0$ limit of ${\rm Re}\left(\partial
  J/\partial \Phi_{s}\right)_{I_b}$ to calculate $K_s$,
so that we can solve Eqs. \eqref{eq:biascurrent} and
\eqref{eq:screeningcurrent} numerically to obtain the working point;
for the moment we assume $Y(0)=0$. 
For the small deviations determining $K_s$, we linearize
Eqs. \eqref{eq:biascurrent} and \eqref{eq:screeningcurrent} and solve
for the real part of the transfer function in the low-frequency limit:
\begin{equation}
{\rm Re}\left(\frac{\partial J}{\partial \Phi_s}\right)_{I_b}=\frac{1}{2 L_j}
\frac{1-\tan^2\Delta\gamma\tan^2\bar{\gamma}}{1+\frac{L}{2 L_j}
(1-\tan^2\Delta\gamma\tan^2\bar{\gamma})}.
\label{eq:real-transfer-function}
\end{equation}
Here, we have introduced the Josephson inductance for one junction, 
$L_j=\Phi_0/2 \pi I_0 \cos\Delta \gamma\cos\bar{\gamma}$.
For $\beta_L \gg 1$,  Eq. \eqref{eq:real-transfer-function} 
approaches $1/L$, while for $\beta_L \ll 1$,
\begin{equation}
{\rm Re}\left(\partial J/\partial \Phi_s\right)_{I_b}=(1/2 L_j)
(1-\tan^2\Delta\gamma\tan^2\bar{\gamma}).
\label{eq:static-coupling-small-B}
\end{equation}
We see that ${\rm Re}\left(\partial J/ \partial \Phi_s\right)_{I_b}$ 
becomes negative for sufficiently high values of 
$I_b$ and $\Phi_s$, which
increase $\bar{\gamma}$ and $\Delta \gamma$.

\begin{figure}
\centering
\includegraphics{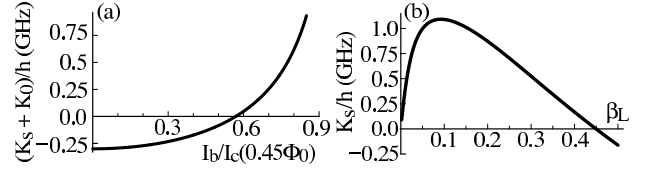}
\caption{(a) Variation of $K$ with $I_b$ for $\Phi_s=0.45 \Phi_0$ and
  device parameters described in text. (b) 
  Highest achievable value of $K_s$ versus $\beta_L$ evaluated
  at $I_b = 0.85 I_c(0.45 \Phi_0)$; $I_0$ (and hence $\beta_L$) 
  is varied for $L=200$ pH.}
\label{coupling-strength}
\end{figure}

We choose the experimentally-accessible SQUID parameters
$L=200$ pH, $C=5$ fF, and $I_0=0.48$ $\mu$A, for which 
$\beta_L = 0.092$. The qubits are characterized by 
$I_q^{(1)}=I_q^{(2)}=0.46$ $\mu$A, $M_{qs}=33$ pH, and $M_{qq}=0.25$
pH, yielding $K_0/h=-0.16$ GHz. Choosing $\Phi_s=0.45 \Phi_0$, 
Eqs.~\eqref{eq:coupling} and \eqref{eq:real-transfer-function} result in
a net coupling strength $K/h=(K_0+K_s)/h$ that is $-0.3$
GHz when $I_b=0$, and zero when $I_b/I_c(0.45 \Phi_0)=0.57$ 
[Fig. \ref{coupling-strength}(a)].
The change in sign of $K_s$ does not occur for all
$\beta_L$.  Figure \ref{coupling-strength}(b) shows
the highest achievable value of $K_s$ versus $\beta_L$. 
We have adopted the optimal design 
at $\beta_L=0.092$.

We also need to consider crosstalk between the coupling and
single-qubit terms in the Hamiltonian. 
When the coupling is switched, in addition to  
$\partial J/\partial \Phi_s$ being altered, 
$J$ also changes, thus shifting the flux biases of the qubits. The calculated 
change in $J$ as the coupler is switched from $I_b=0$
to $I_b/I_c(0.45 \Phi_0)=0.57$
produces a change in the flux in each qubit
corresponding to an energy shift $\delta \epsilon_1/h = \delta \epsilon_2/h =
1.64$ GHz.
In addition, when
the qubits are driven by microwaves to produce single-qubit rotations,
the microwave flux may also couple to $\Phi_s$. As a result,
$K$ is weakly modulated when the coupling would nominally be turned
off.
A typical microwave drive
$\tilde{\epsilon}_i(t)/h$ of amplitude 1 GHz results in a variation of about 
$\pm 14$ MHz 
about $K=0$.

When the bias current is increased to switch off the coupling, the SQUID
symmetry is broken and the qubits are coupled to the noise generated
by the admittance $Y$. 
We estimate the decoherence due to this process by
calculating the environmental spectral density ${\cal J}(\omega)$ in
the spin-boson model  
\cite{Wilhelm03}.
We obtain ${\cal J}(\omega)$ from the classical equation of motion 
for the qubit flux with the 
dissipation from $Y$ coupled to either
qubit through $J$:
\begin{equation}
\mathcal{J}(\omega)=
\left(I_{q}^2 M_{qs}^2/h\right){\rm Im}\left(\partial J/\partial 
\Phi_s\right)_{I_b}.
\end{equation}
To calculate ${\cal J}(\omega)$, we linearize Eqs. \eqref{eq:biascurrent} and
\eqref{eq:screeningcurrent}
around the equilibrium point 
to obtain
\begin{equation}
d\bar{\gamma}=\frac{2\tan\bar{\gamma}\tan\Delta\gamma}{L_j}
\frac{1}{2/L_j-2\omega^2C+i\omega Y}d\Delta\gamma.
\end{equation}
For the case $Y^{-1}=R$, following the path to the static transfer function 
Eq.~\eqref{eq:real-transfer-function} 
and taking the imaginary part in the low-$\beta_L$ limit, 
we obtain
${\rm Im} (\partial J/ \partial\Phi_s)_{I_b}=-\omega/{\cal R}=(\omega/4
  R)\tan^2\Delta\gamma\tan^2\bar{\gamma}$. 
Thus ${\cal J}(\omega)=\alpha\omega$, where
$\alpha=(M_{qs}^2I_q^2/4 h R)\tan^2\Delta\gamma\tan^2\bar{\gamma}$,
and $\alpha(I_b=0)=0$. As $I_b$ is increased to change the coupling strength, 
$\alpha$
increases monotonically.  
For the parameters described above and for $R=2.4$ k$\Omega$, when the
net coupling is zero [$I_b/I_c(0.45 \Phi_0)=0.57$,  
Fig. \ref{coupling-strength}(a)] we find $\alpha = 8 \times 10^{-5}$,
corresponding to a qubit dephasing time of about $500$ ns, one order
of magnitude larger than values currently measured in flux 
qubits \cite{Chiorescu03}.

We now show that this configuration 
implements universal quantum logic efficiently.  
Any $n$-qubit quantum operation can be decomposed into combinations of
two-qubit entangling gates, for example, CNOT, and single-qubit gates
~\cite{Barenco95}.
Single-qubit gates generate local unitary transformations in the
complex 2-dimensional subspace for the corresponding individual qubit,
while the two-qubit gates correspond to unitary transformations in the
4-dimensional Hilbert space. Two-qubit gates which cannot be
decomposed into a product of single-qubit gates are said to be
nonlocal, and may lead to entanglement between the two
qubits~\cite{Zhang03}.
Since we can adjust the qubit coupling $K$ to zero,  
we can readily
implement single-qubit gates with microwave pulses as described below.

To implement the nonlocal two-qubit CNOT gate, we use the concept of
local equivalence: the two-qubit gates $U_1$ and $U_2$ are locally
equivalent if $U_1=k_1U_2k_2$, where $k_1$ and $k_2$ are local
two-qubit gates which are combinations of single-qubit gates applied
simultaneously. These unitary transformations on the
two single-qubit subspaces transform the gate $U_2$ into $U_1$.
The local gate which precedes $U_2$, $k_2$, is given by $k_{21}\otimes
k_{22}$, where $k_{21(22)}$ is a single-qubit gate for qubit $1(2)$,
while the local gate which follows $U_2$, $k_1$, is $k_{11}\otimes
k_{12}$, where $k_{11(12)}$ is a single-qubit gate for
qubit $1(2)$~\cite{Makhlin02}.
Our strategy is to find efficient implementation of a nonlocal quantum
gate $U_2$ that differs only by local gates, $k_1$ and $k_2$, from
CNOT, using the methods in~\cite{Zhang03}, and then to add those local
operations required to achieve a CNOT gate in the computational basis,
in which the SQUID measures the projection of each qubit state vector
onto the z-axis.

The local equivalence classes of two-qubit operations 
have been shown \cite{Zhang03} to be in one-to-one correspondence 
with points in a
tetrahedron, the Weyl chamber.  In this geometric
representation, any two-qubit operation is associated with 
the point $[c_1, c_2, c_3]$, where CNOT corresponds to $[\pi/2,0,0]$.
Furthermore, the nonlocal two-qubit gates
generated by a Hamiltonian acting for time $t$ can be mapped to a
trajectory in this space~\cite{Zhang03}.  
If $K$ is increased instantaneously to a constant value, 
the trajectory generated by 
Eq.~\eqref{eq:hamiltonian}
is well described by the following periodic curve
\begin{equation}
[c_1,c_2,c_3] = \left[K v t/\hbar,p \left|\sin \omega t\right|,
p\left|\sin \omega t\right|\right]. \label{eq:c-two-three}
\end{equation}
Here, $p$ is a function of the system parameters,
$v=\epsilon_1^0\epsilon_2^0/\Delta E_1 \Delta E_2$, and 
$\omega=(\Delta E_1-\Delta E_2)/2\hbar$, where $\Delta 
E_i=[(\epsilon_i^0)^2+\delta_i^2]^{1/2}$ is
the single-qubit energy level splitting. Independently of
$p$, this trajectory reaches $[\pi/2,0,0]$
in a time $t_{K}=n\pi/\omega$ when the
coupling strength is tuned to $K=\hbar \omega/2 n v$, with $n$ a nonzero
integer.

\begin{figure}
\centering
\includegraphics[width=3in]{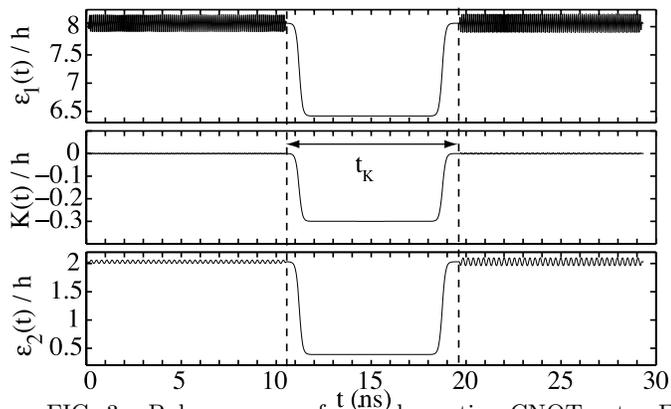}
\caption{
  Pulse sequence for implementing CNOT gate. Energy scales in GHz. 
  Total single-qubit energy bias
  $\epsilon_i(t)=\epsilon_i^0+\tilde{\epsilon}_i(t)+\delta \epsilon_i(t)$, 
  where microwave pulses
  $\tilde{\epsilon}_{1,2}(t)$
  produce single-qubit rotations in the decoupled configuration; 
  crosstalk modulation of $K(t)$ is shown (see text). The bias
  current is pulsed to turn on the interaction in the central region.}
\label{pulse-sequence}
\end{figure}

While this 
analytic solution contains the essential physics, it is
an approximation and does not include vital experimental features, 
in particular, crosstalk and the finite rise time of the bias current pulse.  
To improve the accuracy, we perform a numerical optimization using 
Eq. \eqref{eq:c-two-three} as a starting point, then
add these corrections.  We use tunnel
frequencies $\delta_1/h=5$\,GHz and $\delta_2/h=3$\,GHz,
and include the shifts of the single-qubit energy biases due to the
crosstalk with $K_s$ in Eq.~\eqref{eq:c-two-three} by adding a shift 
$\delta \epsilon_i$ proportional to $K$.
We account for the rise and fall times of the current pulse by
using pulse edges with $90\%$ widths of $0.5$\,ns
[see $K(t)$ in Fig.~\ref{pulse-sequence}].  We numerically
optimize the variable
parameters to minimize the Euclidean distance between the actual
achieved gate and the desired Weyl chamber target CNOT gate.  
We find $K/h=-0.30$\,GHz,
$\epsilon_1^0/h=8.06$\,GHz, $\epsilon_2^0/h=2.03$\,GHz, and
$t_{K}=8.74$\,ns; $t_K$ is the time during which the
qubit coupling is turned on.

As outlined above, to achieve a true CNOT gate we still have to
determine the pulse sequences which implement the 
requisite local gates 
that take this Weyl chamber target $U_2$ to CNOT in the computational
basis. 
Local gates may be implemented by applying microwave radiation, 
$\tilde{\epsilon}_i(t)$, which couples to $\sigma_z^{(i)}$, and is at or near
resonance with the single-qubit energy level splitting $\Delta E_i$.
We note that the single-qubit Hamiltonian of
Eq.~\eqref{eq:single_qubit} driven by a resonant oscillating microwave
field does not permit one to use standard NMR pulses,
since the static and oscillating fields are not
perpendicular, but rather are canted by an angle $\tan
^{-1}(\delta_i/\epsilon_i^0)$. 
To simplify the pulse sequence, we keep $\epsilon_{1,2}^0$ constant at
the values used for the non-local gate generation.
This imposes an additional constraint on the local gates: 
to generate 
a local two-qubit gate $k_1=k_{11}\otimes k_{12}$, the two
single-qubit gates $k_{11}$ and $k_{12}$ must be simultaneous and of
equal duration.  
We satisfy this constraint by making the microwave pulse addressing
one qubit resonant and that
addressing the other slightly off-resonance. Using this
offset
and the relative amplitude and phase of the two microwave pulses 
as variables, we can achieve two different single-qubit gates 
simultaneously, leading to our required local two-qubit gate.

The resulting pulse sequences for $K$ and $\tilde{\epsilon}_{1,2}$ 
are shown in Fig.~\ref{pulse-sequence}. 
The gate has a maximum deviation from CNOT in the
computational basis of $1.6\%$ in any matrix element.
This error arises predominantly from the cross-coupling of the
microwave signals for the two qubits and  
the weak modulation of the $K=0$ state of the coupler 
during the single-qubit microwave manipulations.
While small, this error could be reduced further by performing the
numerical optimization with higher precision 
or by coupling the microwave flux selectively to each of the qubits
and not to the SQUID.
The total elapsed time of $29.35$\,ns is
comparable to measured dephasing 
times in a single flux qubit \cite{Chiorescu03}.

\begin{figure}[t]
\centering
\includegraphics[width=2.5in]{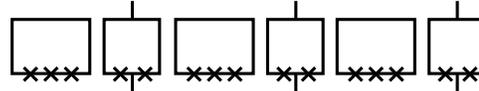}
\caption{
  Chain of flux qubits with intervening dc SQUIDs arranged to provide
  both variable nearest neighbor coupling and qubit readout.}
\label{many-qubits}
\end{figure}

In summary, we have shown that the inverse dynamic inductance of a 
dc SQUID with low $\beta_L$ in the zero-voltage state can be varied by pulsing 
the bias current. 
This technique provides a
variable-strength interaction $K_s$ between flux qubits coupled to
the SQUID, and enables cancellation of the direct mutual inductive
coupling $K_0$ between the qubits so that the net coupling $K$ can be switched 
from a
substantial value to zero. 
By steering a nonlocal gate
trajectory
and combining it with local
gates composed of simultaneous single-qubit rotations driven by
resonant and off-resonant microwave pulses, we have shown that a
simple pulse sequence containing a single switching of the flux
coupling 
for fixed static flux biases results in a CNOT
gate and full entanglement of two flux qubits on a timescale
comparable to
measured decoherence times for flux qubits.
Furthermore, the same SQUID can be used 
to determine the flux state of the qubits. This approach should be
readily scalable to
larger numbers of qubits, as, for example, in Fig. \ref{many-qubits}.

This work was supported by the Air Force Office of Scientific
Research under Grant F49-620-02-1-0295, the Army Research Office under
Grants DAAD-19-02-1-0187 and P-43385-PH-QC, and the National 
Science Foundation under Grant EIA-020-5641. FKW acknowledges travel
support from DFG within SFB 631.

\end{document}